\documentclass[a4paper,fleqn,usenatbib]{mnras}

\usepackage[T1]{fontenc}
\usepackage{ae,aecompl}

\usepackage{graphicx}	% Including figure files
\usepackage{amsmath}	% Advanced maths commands
\usepackage{amssymb}	% Extra maths symbols

\title[LAT search for gamma-ray emission from SF galaxies]{Search for gamma-ray emission from star-forming galaxies with \emph{Fermi} LAT}

\author[Rojas-Bravo, Araya]{
C\'esar Rojas-Bravo,$^{1}$\thanks{E-mail: cesar.rojasbravo@ucr.ac.cr}
Miguel Araya$^{1}$ 
\\
$^{1}$Escuela de F\'isica, Universidad de Costa Rica
}

\date{Accepted XXX. Received YYY; in original form ZZZ}

\pubyear{2016}

\begin{document}
\label{firstpage}
\pagerange{\pageref{firstpage}--\pageref{lastpage}}
\maketitle

\begin{abstract}
Recent studies have found a positive correlation between the star-formation rate of galaxies and their gamma-ray luminosity. Galaxies with a high star-formation rate are expected to produce a large amount of high-energy cosmic rays, which emit gamma-rays when interacting with the interstellar medium and radiation fields. We search for gamma-ray emission from a sample of galaxies within and beyond the Local Group with data from the LAT instrument onboard the \emph{Fermi} satellite. We exclude recently detected galaxies (NGC 253, M82, NGC 4945, NGC 1068, NGC 2146, Arp 220) and use seven years of cumulative ``Pass 8'' data from the LAT in the 100 MeV to 100 GeV range. No new detections are seen in the data and upper limits for the gamma-ray fluxes are calculated. The correlation between gamma-ray luminosity and infrared luminosity for galaxies obtained using our new upper limits is in agreement with a previously published correlation, but the new upper limits imply that some galaxies are not as efficient gamma-ray emitters as previously thought.

\end{abstract}
%instruction for authors:http://www.oxfordjournals.org/our_journals/mnras/for_authors/

\section{Introduction}
\label{sec:intro}
Starburst galaxies have a high star-formation rate (SFR) and are known to emit at GeV and TeV energies \citep{Acero2009,Abdo2010a,lenain2011,Ackermann2012}. Diffuse high-energy and very high-energy gamma rays are tracers of non-thermal processes resulting from the propagation and interaction of cosmic rays with radiation and magnetic fields, and the high-star formation rates are expected to be correlated with the abundance of particle accelerators, such as supernova remnants and pulsar wind nebulae.

Containing dense gas regions and active star-formation, starburst galaxies are suitable objects to study the origin and physics of cosmic rays and their interactions with the interstellar medium \citep[e.g.,][]{Ohm2015}, and may also contribute as a population to the isotropic diffuse gamma-ray emission \cite[e.g.,][]{Pavlidou2002}. Cosmic rays can produce gamma-ray emission mainly via two mechanisms: inelastic collisions between cosmic ray nuclei and the interstellar gas producing, among other particles, neutral pions, which themselves decay into gamma rays, and gamma-ray emission from inverse Compton scattering and bremsstrahlung processes from high-energy electrons \citep{Dermer1986,Ohm2015,Drury1994}.

The interstellar gas efficiently absorbs star emission and re-emits it in the infrared (IR). IR emission dominates the spectral energy distribution of actively star-forming galaxies, which often have huge IR luminosities \citep{Sanders1996}, and it is a good indicator of star formation rates. IR emission is also observed to be correlated with radio continuum emission \citep{Condon1991} which is produced by high-energy electrons and positrons as they spiral in the interstellar magnetic fields. It is therefore also expected that a correlation exists between IR emission and diffuse gamma-ray emission, unless particle acceleration takes place a long time after a star formation period, which is thought not to be the case (a SNR is thought to be able to accelerate particles for several $\sim10^4$ years, much less than the lifetimes of massive stars).

Recently, detections of high-energy emission from starburst galaxies such as M82 and NGC 253 by gamma-ray detectors, both space-based and imaging air-Cerenkov telescopes \citep{Abdo2010a,Acciari2009,Acero2009} have been made, and were already predicted to emit gamma-rays by some authors due to their high supernova explosion rate and dense gas content \citep{Voelk1989,Paglione1996,Persic2008,Pavlidou2002,Torres2004,Thompson2007,Stecker2007,Persic2010,Lacki2011}. Additionally, the Milky Way Galaxy \citep{Strong2010}, the Large \citep{Abdo2010b} and Small \citep{Abdo2010c} Magellanic Clouds and M31 \citep{Abdo2010d} are known to have a diffuse gamma-ray emission component at GeV energies. We refer the reader to the study by \citet{Ackermann2012} for further details on some of the known photometric estimators of star forming activity.

An unprecedented long-term survey of the sky at GeV energies, with ever increasing statistics, has been carried out by the \emph{Fermi} satellite (see \cite{Atwood2009} for instrument details). Using three years of ``Pass 7'' data from the LAT instrument onboard this satellite, \citet{Ackermann2012} examined a sample of 64 star-forming galaxies beyond the Local Group, selected on the basis of their mid-high SFR, to search for gamma-ray emission. They found that a simple power law correlation between the gamma-ray luminosity and SFR holds true in the sample. This correlation is the same that had been previously observed for the galaxies in the Local Group \citep{Abdo2010b, Abdo2010c,Abdo2010d}. The galaxies NGC 4945 and NGC 1068 were added to the group of firm high-energy detections. Later, some other galaxies have been found at GeV energies, including Circinus \citep{Hayashida2013}, NGC 2146 \citep{Tang2014} and, very recently, a tentative detection for Arp 220 has been claimed \citep{Peng2016,Griffin2016}. Note that galaxies which contain an active galactic nucleus (AGN) could also produce nonthermal radio and gamma-ray emission by a mechanism that is unlikely that revealed by the SFR-gamma-ray correlation.

In this work, we present the results of a LAT analysis with more statistics, including seven years of cumulative data with better instrument performance (``Pass 8'') for the galaxy sample studied by \citet{Ackermann2012}. No new detections were made and upper limits on the fluxes were found for all galaxies. In Section \ref{sec:data} we present the data analysis and the results are shown in Section \ref{sec:results}. We give a discussion of our results in the context of the known correlation between infrared and gamma-ray emission in Section \ref{sec:discussion} which we compare to the new correlation resulting from our data.

\section{Data}\label{sec:data}
\subsection{Galaxy sample}\label{sec:galsample}
The set of galaxies selected for analysis was that as given in Table 1 of \citet{Ackermann2012}. These galaxies were chosen based on the HCN survey of \citet{Gao2004}. The HCN $J$ = 1-0 line emission is stimulated in dense molecular gas which is associated with molecular clouds where most of the star formation occurs (see \citet{Ackermann2012} for additional details). The sample also excludes galaxies with Galactic longitude $|b|<10^{\circ}$, to avoid contamination from diffuse gamma-rays near the plane of the Milky Way. The largest redshift of the galaxies in the sample is $\sim0.06$, and most distances were calculated with $H_0=75$ km s$^{-1}$ Mpc $^{-1}$ except for a few nearby galaxies, for which the mostly used and recent value in the literature is adopted \citep{Gao2004}.

From  Ackermann's original 64-galaxy list, we excluded from the analysis those that were already detected, namely NGC 253, M82, NGC 4945, NGC 1068 \citep{Ackermann2012}, NGC 2146 \citep{Tang2014} and Arp 220 \citep{Peng2016,Griffin2016}. We also excluded Local Group galaxies which have been seen at GeV energies (LMC, SMC, M31) and added M33, which has not been seen. The final galaxy sample for which we analyzed LAT data therefore consisted of 59 galaxies. They are classified, according to SIMBAD's database\footnote{see http://simbad.u-strasbg.fr/simbad/}\citep{Wenger2000} as starburst galaxies, Seyfert galaxies, HII galaxies and LINER galaxies, among others.

\citet{Ackermann2012} used the Swift BAT 58 month survey catalog with AGN-classifications to determine which of the galaxies in the sample contain an AGN. According to this catalog, only 7 out of the 59 galaxies in the sample host an AGN. On the other hand, SIMBAD's database lists 25 of the galaxies in the same sample as hosting an AGN. Not all AGN emit gamma-rays. For example it has been found that Seyfert galaxies hosting radio-quiet AGN are generally gamma-ray quiet (e.g., \citet{Teng2011}). Throughout the rest of this work we follow SIMBAD's classification scheme.

\subsection{Data analysis}\label{sec:dataan} 
We collected \emph{Fermi}-LAT data in sky-survey mode from 2008 August 7 to 2015 August 20 (mission elapsed times between 239557420 and 461682600 s). We used the most recent photon and spacecraft data release (PASS 8), which provides improved event reconstructions, an increased effective area and better energy measurements \footnote{see http://fermi.gsfc.nasa.gov/ssc} with respect to previous releases, and the instrument response functions P8R2\_SOURCE\_V6P.

Data were analyzed with the public Science Tools software for LAT data, version v10r0p5, with the recommended cuts applied, including selecting zenith angles less than 90$^{\circ}$ to remove background photons and keeping SOURCE class events. We selected a 20$^{\circ}\times20^{\circ}$ region of interest (RoI) around the catalogued position of each galaxy, and included events with energies between 100 MeV and 100 GeV. Using a large RoI is necessary to account for the broad PSF of the instrument, particularly at the lowest energies. We binned the data in counts maps with a scale of 0.1$^{\circ}$ per pixel and used 30 logarithmically spaced bins in energy. Time intervals are chosen when the LAT was in science operations mode and data quality is good.

The spectral and morphological properties of LAT sources were obtained through a maximum likelihood analysis\footnote{see http://fermi.gsfc.nasa.gov/ssc/data/analysis/\\scitools/binned\_likelihood\_tutorial.html}, as it is common in modeling of gamma-ray data \citep{Mattox1996}. The spectral parameters and morphology of the sources in the model were adjusted to maximize the likelihood of the fit. The analysis included all sources from the most recent \emph{Fermi} LAT catalog (3FGL) as well as the standard galactic diffuse emission model, given by the file gll\_iem\_v06.fits, and the isotropic component, described by the file iso\_P8R2\_SOURCE\_V6\_v06.txt, which takes into account the extragalactic background and misclassified cosmic rays. These models are distributed with the LAT analysis software.

During the maximum likelihood fitting, the normalizations of the diffuse components and the spectral normalizations of cataloged LAT sources located within 9$^{\circ}$ of the center of the RoI were kept free. After the fit, the residuals maps were obtained by subtracting the best-fit model from the data. This was also done at higher energies for comparison (above 200 MeV, 500 MeV and 1 GeV) whenever the regions showed substantial residuals or there were nearby sources or considerable Galactic background emission. In most cases, though, the residuals above 100 MeV did not show the characteristic emission from a point source, as it would be expected for an extragalactic object such as a normal galaxy.

As a next step, we carried out the fit again in the corresponding RoI after incorporating a point source at the position of each of the galaxies in the sample to evaluate its significance. A simple power-law spectrum was used for each galaxy, of the form $ dN / dE \propto E ^{ - \Gamma}$, where $\Gamma$ is the spectral index. The significance can be estimated with the test statistic (TS), which is one of the main parameters that result from the likelihood fit. It is defined as $-2$ log($L_0/L$),where $L_0$ is the maximum likelihood value for a model without an additional source (the ``null hypothesis'') and $L$ is the maximum likelihood value for a model including the additional source at a specified location. The resulting source significance can be estimated as $\sqrt{\mbox{TS}}$ \citep{Mattox1996} and therefore a tentative detection is made if the TS is greater than 25. 
\section{Results}\label{sec:results}
\subsection{LAT Data Analysis Results}\label{sec:latanalysis}
Results from the maximum likelihood analysis for each source are shown in Table \ref{tab1}. No galaxies were significantly detected. Upper limits on the flux, calculated at the 95\% confidence level are shown with the corresponding upper limits on gamma-ray luminosities. The distance and infrared luminosity are also shown and were taken from \cite{Ackermann2012}. Upper limits are obtained assuming a source photon index of $\Gamma = 2.2$, which is typical for detected star-forming galaxies. All flux upper limits are lower than the values, based on a smaller data set and PASS 7 analysis, reported by \cite{Ackermann2012} - except for IC 342 and NGC 5135.

A group of sources apparently have TS > 20, but their fitted spectral indices are not reasonable from a physical point of view. A point source at the position of IC 342, for example, gives a TS of 91. Several issues with the data or the analysis could cause an artificially high TS value: e.g., the presence of bright nearby sources or a diffuse background emission that is not accurately modelled. After redoing the fits with a fixed spectral index of 2.2, the TS values drop below the detection threshold for all sources that had showed an apparently high significance. To double check that the sources were not detected, we carried out the likelihood fits at higher energies, namely above 200 MeV and above 500 MeV to lower the uncertainties introduced by different factors. We confirm there are no new detections. The results for this small group of sources with high artificial TS values are shown in Table \ref{tab2}. Above 200 MeV, the TS of all the sources drop substantially. Above 500 MeV, the source that shows the highest TS is NGC 1530 (8.6).

In some RoIs there are catalogued gamma-ray sources near the position of some of the galaxies, such as NGC 6701, which is about 0.7$^{\circ}$ from the unidentified source 3FGL J1840.5+6116. Other galaxies that are located relatively near LAT sources are NGC 4826 (located 0.7$^{\circ}$ from 3FGL J1254.5+2210), NGC 1365 (located 0.7$^{\circ}$ from 3FGL J0336.9-3622), NGC 2276 (located 0.7$^{\circ}$ from 3FGL J0746.9+8511), NGC 4041 (located 0.9$^{\circ}$ from 3FGL J1155.9+6136) and Arp 55 (located $\sim 1^{\circ}$ from 3FGL J0920.9+4442). There is a LAT source candidate located 0.5$^{\circ}$ from NGC 4826 (3FGL J1258.4+2123), with no association and it is reported with a significance of 5$\sigma$ in the latest LAT catalog. The rest of LAT sources are gamma-ray counterparts of AGN. From their locations or associations we conclude that it is very unlikely that any of these LAT sources are the counterparts of star-forming galaxies.

Next, we discuss two objects for which a particularly careful analysis in order to calculate the upper limit was necessary.

\begin{table*}
	\begin{center}
	\caption{Upper limits (95\% CL) on gamma-ray flux and luminosity for star-forming galaxies with PASS 8 LAT data}
	\label{tab1}
	\begin{tabular}{lccccc} 
		\hline
		Galaxy & $D$ & $L_{8-1000  \,\micron}$ & $F_{0.1-100 \,\mbox{\tiny GeV}}$ & $\Gamma$ & $L_{0.1-100 \,\mbox{\tiny GeV}}$ \\
		& (Mpc) & (10$^{10}$ $L_{\sun}$) & (10$^{-9}$ ph cm$^{-2}$ s$^{-1}$) &  & (10$^{40}$ erg s$^{-1}$)  \\
		\hline
		M33 & 0.85 & 1.2 & $<$3.2 & 2.2 & 0.02\\
		IC 342 & 3.7 & 1.4 & $<$2.9 & 2.2 & 0.3\\
		M83 & 3.7 & 1.4 & $<$1.1 & 2.2 & 0.1\\
		NGC 4826 & 4.7 & 0.26 & $<$1.0 & 2.2 & 0.2\\
		NGC 6946 & 5.5 & 1.6 & $<$0.6 & 2.2 & 0.2\\
		NGC 2903 & 6.2 & 0.83 & $<$1.5 & 2.2 & 0.5\\
		NGC 5055 & 7.3 & 1.1 & $<$0.9 & 2.2 & 0.4\\
		NGC 3628 & 7.6 & 1.0 & $<$1.3 & 2.2 & 0.6\\
		NGC 3627 & 7.6 & 1.3 & $<$1.4 & 2.2 & 0.7\\
		NGC 4631 & 8.1 & 2.0 & $<$0.9 & 2.2 & 0.5\\
		NGC 4414 & 9.3 & 0.81 & $<$0.7 & 2.2 & 0.6\\
		M51 & 9.6 & 4.2 & $<$0.9 & 2.2 & 0.8\\
		NGC 891 & 10.3 & 2.6 & $<$2.9 & 2.2 & 2.6\\
		NGC 3556 & 10.6 & 1.4 & $<$1.6 & 2.2 & 1.5\\
		NGC 3893 & 13.9 & 1.2 & $<$0.6 & 2.2& 1.0\\
		NGC 660 & 14.0 & 3.7 & $<$1.5 & 2.2 & 2.5\\
		NGC 5005 & 14.0 & 1.4 & $<$1.5 & 2.2 & 2.6\\
		NGC 1055 & 14.8 & 2.1 & $<$2.2 & 2.2 & 4.1\\
		NGC 7331 & 15.0 & 3.5 & $<$1.6 & 2.2 & 3.2\\
		NGC 3079 & 16.2 & 4.3 & $<$2.0 & 2.2 & 4.4\\
		NGC 4030 & 17.1 & 2.1 & $<$0.6 & 2.2 & 1.4\\
		NGC 4041 & 18.0 & 1.7 & $<$1.2 & 2.2 & 3.4\\
		NGC 1365 & 20.8 & 13 & $<$1.8 & 2.2 & 6.9\\
		NGC 1022 & 21.1 & 2.6 & $<$1.6 & 2.2 & 6.0\\
		NGC 5775 & 21.3 & 3.8 & $<$0.9 & 2.2 & 3.7\\
		NGC 5713 & 24.0 & 4.2 & $<$0.3 & 2.2 & 1.4\\
		NGC 5678 & 27.8 & 3.0 & $<$0.7 & 2.2 & 4.6\\
		NGC 520 & 31.1 & 8.5 & $<$0.6 & 2.2 & 4.6\\
		NGC 7479 & 35.2 & 7.4 & $<$2.7 & 2.2 & 29.0\\
		NGC 1530 & 35.4 & 4.7 & $<$2.5 & 2.2 & 26.6\\
		NGC 2276 & 35.5 & 6.2 & $<$0.7 & 2.2 & 7.9\\
		NGC 3147 & 39.5 & 6.2 & $<$0.3 & 2.2 & 3.6\\
		Arp 299 & 43.0 & 63 & $<$1.5 & 2.2 & 23.9\\
		IC 5179 & 46.2 & 14 & $<$0.4 & 2.2 & 7.5\\
		NGC 5135 & 51.7 & 14 & $<$1.8 & 2.2 & 41.2\\
		NGC 6701 & 56.8 & 11 & $<$0.6 & 2.2 & 16.5\\
		NGC 7771 & 60.4 & 21 & $<$1.3 & 2.2 & 39.6\\
		NGC 1614 & 63.2 & 39 & $<$1.6 & 2.2 & 54.4\\
		NGC 7130 & 65.0 & 21 & $<$0.3 & 2.2 & 11.5\\
		NGC 7469 & 67.5 & 41 & $<$1.2 & 2.2 & 46.3\\
		IRAS 18293-3413 & 72.1 & 54 & $<$0.3 & 2.2 & 13.7\\
		Mrk 331 & 75.3 & 27 & $<$0.4 & 2.2 & 20.6\\
		NGC 828 & 75.4 & 22 & $<$1.0 & 2.2 & 48.8\\
		IC 1623 & 81.7 & 47 & $<$0.7 & 2.2 & 38.5\\
		Arp 193 & 92.7 & 37 & $<$1.4 & 2.2 & 101.4\\
		NGC 6240 & 98.1 & 61 & $<$0.4 & 2.2 & 31.1\\
		NGC 1144 & 117.3 & 25 & $<$0.5 & 2.2 & 54.1\\
		Mrk 1027 & 123.5 & 26 & $<$0.7 & 2.2 & 95.4\\
		NGC 695 & 133.5 & 47 & $<$1.9 & 2.2 & 291.7\\
		Arp 148 & 143.3 & 36 & $<$1.1 & 2.2 & 196.4\\
		Mrk 273 & 152.2 & 130 & $<$0.6 & 2.2 & 124.9\\
		UGC 05101 & 160.2 & 89 & $<$0.5 & 2.2 & 117.8\\
		Arp 55 & 162.7 & 46 & $<$0.8 & 2.2 & 185.2\\
		Mrk 231 & 170.3 & 300 & $<$0.9 & 2.2 & 226.4\\
		IRAS 05189-2524 & 170.3 & 120 & $<$0.5 & 2.2 & 121.2\\
		IRAS 17208-0014 & 173.1  & 230 & $<$1.4 & 2.2 & 351.1\\
		IRAS 10566+2448 & 173.3 & 94 & $<$0.4 & 2.2 & 91.3\\
		VII Zw 31 & 223.4 & 87 & $<$0.2 & 2.2 & 90.3\\
		IRAS 23365+3604 & 266.1 & 140 & $<$0.9 & 2.2 & 568.5\\
		
		\hline
        \end{tabular}\\
\end{center}
\footnotesize{Galaxy distances ($D$) and total IR ($8-1000\,\micron$) luminosities are shown here also \citep[see][]{Ackermann2012}. $\Gamma$ is the spectral index used for the upper limit calculation.}
\end{table*}

\begin{table*}
	\centering
	\caption{Likelihood results for galaxies with apparent high TS values for different energy intervals, using free and fixed spectral index}
	\label{tab2}
	\begin{tabular}{lccccc} 
		\hline
		Galaxy & TS & $\Gamma$\textsuperscript{$\dagger$} & TS& TS & TS  \\
		& above 100 MeV &above 100 MeV& above 100 MeV & above 200 MeV & above 500 MeV   \\
		& $\Gamma$ free &  & $\Gamma$ fixed at 2.2& $\Gamma$ fixed at 2.2 & $\Gamma$ fixed at 2.2  \\
		\hline
		IC 342  & 91.0 &6.3$\pm0.6$& 4.6 & 1.9 & 2.9  \\
		NGC 4826 & 25.1 &5.1$\pm1.2$& 0.1 & 2.6 & 2.1  \\
		NGC 6946 & 30.3 &6.7$\pm1.1$& 0.0 & 0.7 & 0.7	  \\
		NGC 4041 & 26.2 &7.5$\pm1.1$& 1.0 & 3.2 & 1.4  \\
		NGC 1365 & 23.3 &7.5$\pm1.2$&3.8& 4.9 & 5.6  \\
		NGC 7479  & 20.0 &4.9$\pm1.4$&5.0& 4.4 & 1.5  \\
		NGC 1530 & 53.1 &6.8$\pm0.9$&8.0& 6.5 & 8.6  \\
		NGC 2276 & 35.2 &7.0$\pm1.0$&0.0& 0.2 & 0.1  \\
		NGC 6701 & 24.3 &7.6$\pm0.2$&0.0& 2.8 & 0.0  \\
		Arp 55 & 20.0 &7.5$\pm1.3$& 0.0 & 0.2 & 0,7  \\
		\hline
	\end{tabular}\\
\textsuperscript{$\dagger$}\footnotesize{Spectral index from fitting a point source at the location of each galaxy above 100 MeV with the statistical uncertainties as printed by the likelihood algorithm. The sources are not detected above 200 MeV and therefore these values are artificial, likely resulting from systematic uncertainties in the analysis around 100 MeV.}
\end{table*}

\subsection{NGC 3627}\label{sec:ngc3627}
Originally, a maximum likelihood analysis for a point source at the position of NGC 3627 (a LINER-type AGN, also known as M66) yielded a TS of 67, with a photon index of 2.8$\pm0.1$. However, we noticed the presence of a nearby source in the counts map, which is not present in the latest LAT catalog. In order to constrain the position of this new source, we used the tool \emph{gtfindsrc} for events above 2 GeV to take advantage of the narrower point-spread function of the LAT at higher energies. The result is RA = 169.83$^{\circ}$, Dec = 12.65$^{\circ}$ with an error circle radius of 0.37$^{\circ}$ for the new source. This is consistent with the position of the quasar 4C 12.39. Assuming a power-law spectral shape for the new source at the position of the quasar we obtain a best-fit photon index of 2.7$\pm0.1$ above 100 MeV. When adding this new source to the model and fitting again above 100 MeV, the new TS value for a point source at the position of NGC 3627 is 4.9.

The angular distance between NGC 3627 and the position of the new source found here is $\sim$ 0.36$ ^{\circ}$. The residuals obtained after fitting a model with both sources, above 1 GeV, might indicate that there are contributions from other sources near the detection threshold in the region. We thus cannot conclude without further analysis if we are seeing a region with confused sources, one of them being the star-forming galaxy. We calculated an upper limit for NGC 3627.

\subsection{NGC 3079}\label{sec:ngc3079}
Similarly, a maximum likelihood analysis for NGC 3079 (a Seyfert 2 galaxy), resulted in a TS value of 28. However, this galaxy is near the very bright source 3FGL J0957.6+5523 and such result cannot be regarded as evidence for a detection. In fact, the resulting spectral index for NGC 3079 above 100 MeV ($\sim7$) is not physically reasonable. In order to lower the contamination from the nearby LAT source, we carried out an analysis above 1 GeV. The source NGC 3079 is not seen above this energy and we calculate an upper limit on its flux above this energy and extrapolate the result above 100 MeV to report the final upper limit. We do not include this source in Table \ref{tab2}. While for the other galaxies listed in that table there are several systematic effects which can produce an artificially large TS value, in the case of NGC 3079 we believe we are seeing the influence of the bright source 3FGL J0957.6+5523.

\section{Discussion}\label{sec:discussion}
\subsection{Gamma-ray and Infrared correlation}
\label{sec:grircomp}
The total IR luminosity (8-100 $\mu $m) is a well-established tracer of star-formation rate (SFR) for late-type galaxies (\cite{Kennicutt1998a}, \citet{Ackermann2012} ). A conversion rate that has been proposed (\cite{Kennicutt1998b} is given by 

\begin{equation}
    \frac{\mbox{SFR}}{\mbox{M}_{\sun}\mbox{yr}^{-1}} =\epsilon1.7\times 10^{-10}\frac{L_{8-1000 \, \mu m}}{L_{\sun}}.
	\label{sfr}
\end{equation}

assuming that the thermal emission from dust is a calorimetric measure of the radiation of young stars, and $\epsilon$ depends on the assumed initial mass function.

\begin{figure}
    \includegraphics[width=\columnwidth]{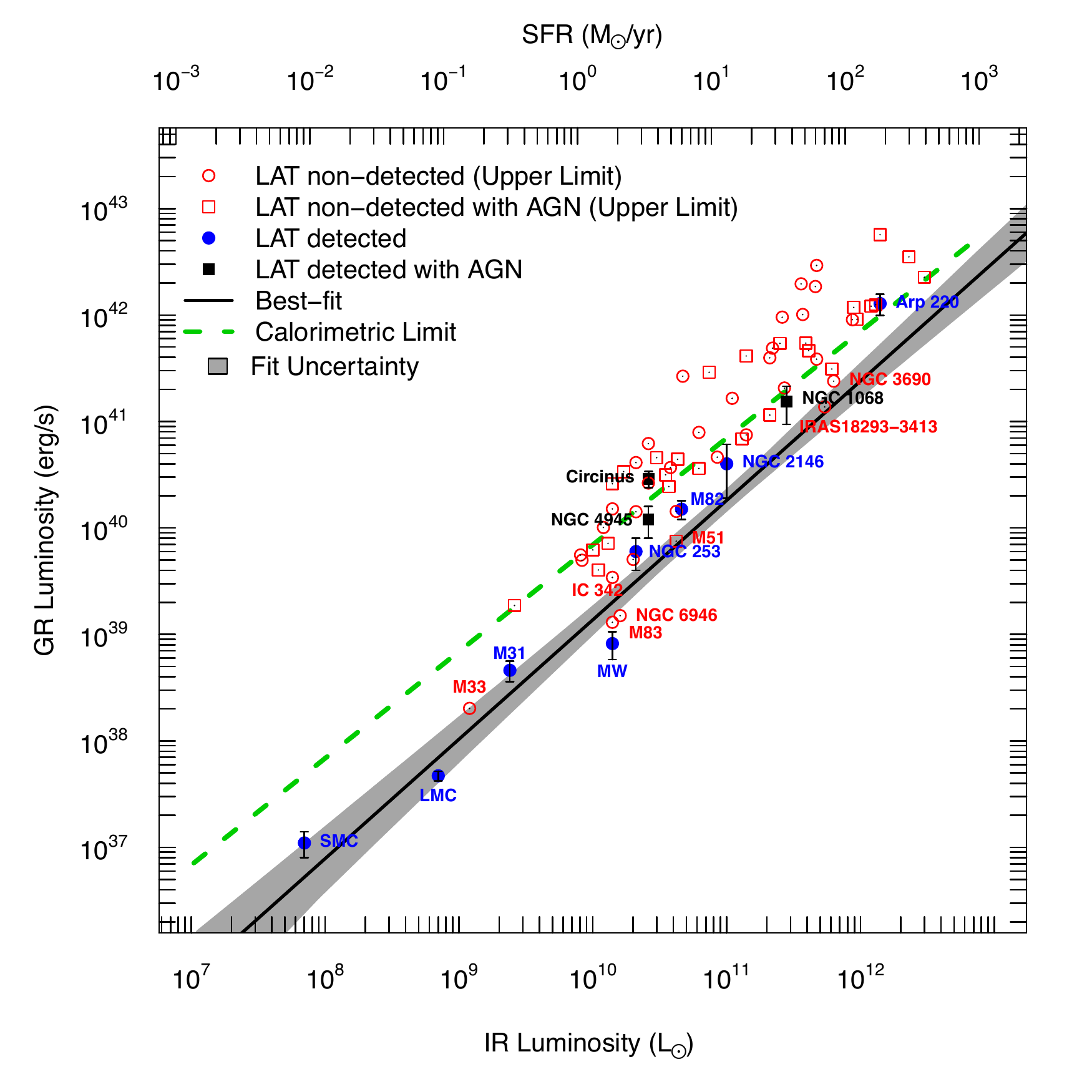}
    \caption{Gamma-ray luminosity (0.1-100 GeV) vs total infrared luminosity (8-1000 $\mu$m) for star-forming galaxies. Upper limits on gamma-ray luminosity are obtained at the 95\% confidence level. The best-fit power-law correlation calculated for these data is shown with its 95\% confidence level uncertainty. Infrared luminosity uncertainties are not shown for clarity.}
    \label{fig1}
\end{figure}

Figure \ref{fig1} compares the gamma-ray luminosity to the total infrared (8-1000 $\mu$m) luminosity for the galaxies in our sample, including the gamma-ray luminosity upper limits obtained in this work. The SFR, estimated from Eq. \ref{sfr} with $\epsilon = 0.79$ \citep{Ackermann2012}, is also shown in the plot. Non-AGN gamma-ray detections by previous authors are indicated by solid, blue circles, while detections of galaxies hosting AGN by solid, black squares. LAT upper limits for galaxies without an AGN are shown as red, hollow circles, while LAT upper limits for galaxies containing an AGN are shown in red, hollow squares.

In order to obtain the best-fit relation between the gamma-ray and infrared luminosities with our data, we applied a regression algorithm known as expectation-maximization, which is commonly used in survival analysis with censored data, such as those including upper limits, as implemented in the code ASURV Rev 1.2 \citep{lavalley1992}. We fitted a linear relation in log-space of the form

\begin{equation}
    \mbox{log}\left(\frac{L_{0.1-100 \,\mbox{\tiny GeV}}}{\mbox{erg s}^{-1}}\right) = m \cdot \mbox{log}\left(\frac{L_{8-1000 \, \mu m}}{L_{\sun}}\right) + b,
	\label{fit}
\end{equation}
and obtained the values for the slope and intercept, $m=1.12\pm 0.08$ and $b=27.9\pm 0.8$, respectively. We note that the uncertainties do not take into account the measurement errors in the gamma-ray fluxes. The best-fit line is shown in Fig. \ref{fig1} and its 95\% confidence level region is also indicated. We compared this fit with that obtained with the Buckley-James algorithm \citep{Buckley1979,Jin2006} and the results are consistent. The Buckley-James method is a generalization for cases in which the distribution of residuals relative to the regression line is not normal and it is estimated using the Kaplan-Meier distribution.

We see that a simple power-law scaling relation is consistent with our data, as has been found before with less stringent gamma-ray upper limits. Our power-law index and normalization are similar to those found by \cite{Ackermann2012}. In particular, the power-law index for their best-fit scaling relation is $1.17\pm0.07$. Our results do not change, within the parameter uncertainties, if we exclude from the fit detected galaxies that contain an AGN (Circinus, NGC 1068 and NGC 4945), whose gamma-ray emission might have contributions from the nuclei.

The calorimetric limit is also indicated in Fig. \ref{fig1} by the dashed line. In this limit, the energy losses of cosmic rays are dominated by their interaction with ambient material in the galaxy rather than by escape. The residence time of cosmic rays in a galaxy is then of the order of the cooling time for hadronic interactions, which is approximately inversely proportional to the product of the average gas density and the cross section for inelastic proton-proton interactions, and thus is almost independent of the cosmic ray energy. This limit can be estimated as a function of the galaxy SFR assuming a value for the average cosmic ray luminosity per supernova explosion (10$^{50}$ erg), and ignoring leptonic contributions to the gamma-ray flux as well as contributions from individual sources \citep[for details, see][]{Ackermann2012}.

Another caveat regarding the use of Eq. \ref{sfr} has to do with the fact that the SFR-IR relation might be different for starburst galaxies in different evolution stages \citep{Vega2005}. Eq. \ref{sfr} would be true for young starbursts with optically thick dust clouds that efficiently reprocess UV radiation. For less active starbursts with lower optical depth, a second infrared component originating from the heating of the ISM by UV emission mainly from old stars arises , and a correction to the infrared emission has to be considered \citep{Vega2005, Persic2007}.

Using our scaling relation between gamma-ray and infrared luminosities and Eq. \ref{sfr} we can estimate maximum calorimetric efficiencies for galaxies as a function of their star formation rates. This efficiency is defined as the ratio of the observed gamma-ray luminosity to the one predicted in the ideal calorimetric limit. The scaling relation found here implies that for galaxies with SFR $\sim 1$ M$_{\sun}$yr$^{-1}$ the maximum calorimetric efficiency is $\sim 14-25$\% and for SFR $\sim 10$ M$_{\sun}$yr$^{-1}$, it is $20-30$\%, which is at the lower end of some previously predicted values \citep{Lacki2011,Ackermann2012}.

For some galaxies such as the the Milky Way and others with similar SFRs, such as M83 and NGC 6946, the observed efficiency is relatively low compared to other galaxies. In fact, the measured gamma-ray upper limits for M83 and NGC 6946 are slightly in conflict with the uncertainty in our scaling relation. Future observations may reveal that some galaxies are not as effective as others at transferring energy to cosmic rays or keeping them confined long enough to show gamma-ray emission at the level predicted by the scaling relation, while on the other hand others such as Arp 220 efficiently emit gamma-rays at the calorimetric level. However, as it has been noted, for some galaxies Eq. \ref{sfr} may not hold true due to the effects of their evolutionary phase on the measured infrared luminosity \citep{Persic2007}, which would modify the predicted calorimetric efficiency.

With respect to predictions on possible LAT detections, M83, M33 and NGC 3690 had been predicted to have the highest probability of being detected by the LAT telescope within $\sim 6$ years of observations \citep{Ackermann2012,Lacki2011}, and this has not been the case, while NGC 2146 and Arp 220 have been detected. As can be seen in Fig. \ref{fig1}, the galaxies IRAS 18293-3413 and M51 have moderate to high SFR and our upper limits on their gamma-ray luminosity place them well within the scaling relation uncertainty, which might make them candidates for a future detection with \emph{Fermi}.

\section*{Acknowledgements}
We are grateful to Universidad de Costa Rica and Escuela de F\'isica for financial support. This research has made use of NASA's Astrophysical Data System and the SIMBAD database, operated at CDS, Strasbourg, France.



%\bibliographystyle{mnras}


%%%%%%%%%%%%%%%%%%%%%%%%%%%%%%%%%%%%%%%%%%%%%%%%%%

%%%%%%%%%%%%%%%%% APPENDICES %%%%%%%%%%%%%%%%%%%%%

%\appendix



% Don't change these lines
\bsp	% typesetting comment
\label{lastpage}
\end{document}